\documentclass[aps,prl,preprint,superscriptaddress]{revtex4}

\usepackage[dvips]{graphicx}
\begin{document}

\title{Frequency-dependent spontaneous emission rate from CdSe and CdTe nanocrystals: influence of dark states.}

\author{A. F. van Driel}
\email[Electronic address: ]{a.f.vandriel@phys.uu.nl}
\affiliation{Debye Institute, Utrecht University, P.O. Box 80 000,
3508 TA Utrecht, The Netherlands}

\author{G. Allan}
\author{C. Delerue}
\affiliation{IEMN (CNRS, UMR 8520), D\'epartement ISEN, 41
Boulevard Vauban, 59046 Lille CEDEX, France}

\author{P. Lodahl}
\altaffiliation{Research Center COM, Technical University of
Denmark, Dk-2800 Lyngby, Denmark}
\affiliation{Complex Photonic Systems (COPS), Department of
Science and Technology and MESA+ Research Institute, University of
Twente, PO Box 217, 7500 AE Enschede, The Netherlands}

\author{W. L. Vos}
\altaffiliation{FOM Institute for Atomic and Molecular Physics
(AMOLF), NL-1098 SJ Amsterdam, The Netherlands}
\affiliation{Complex Photonic Systems (COPS), Department of
Science and Technology and MESA+ Research Institute, University of
Twente, PO Box 217, 7500 AE Enschede, The Netherlands}

\author{D. Vanmaekelbergh}
\affiliation{Debye Institute, Utrecht University, P.O. Box 80 000,
3508 TA Utrecht, The Netherlands}
\date{\today}

\begin{abstract}
We studied the rate of spontaneous emission from colloidal CdSe
and CdTe nanocrystals at room temperature. The decay rate,
obtained from luminescence decay curves, increases with the
emission frequency in a supra-linear way. This dependence is
explained by the thermal occupation of dark exciton states at room
temperature, giving rise to a strong attenuation of the rate of
emission. The supra-linear dependence is in agreement with the
results of tight-binding calculations.
\end{abstract}


\maketitle

Semiconductor nanocrystals have a huge potential for application
as monochromatic light-sources in biological research, photonic
studies and opto-electrical
devices\cite{1982Efros,1984Brus,1996Efros,1998Chan,1998Bruchez,2000Michler,2002Coe,2004Wanga,2004Lodahl}.
The most important class of semiconductor nanocrystals are
colloidal nanocrystals, in practice CdTe and CdSe. Due to quantum
confinement, nanocrystals possess discrete electron and hole
energy levels\cite{1982Efros,1984Brus}. As a consequence, optical
absorption occurs at discrete energies which are determined by the
size and the shape of the nanocrystal host\cite{1996Efros}. At low
excitation density, light emission is due to decay of the lowest
exciton state to the ground state. The possibility of tailoring
the exciton emission energy by the nanocrystal size has led to a
world-wide interest in light-emitting semiconductor nanocrystals.
An open question is how the size and the emission frequency of a
semiconductor nanocrystal control the excitons' spontaneous
emission decay rate. Such a study is currently feasible due to the
availability of colloidal nanocrystal suspensions with a high
quality. These suspensions form a unique model system to probe the
frequency dependence of Fermi's golden rule; the emission
frequency can be tuned via the crystal size without changing the
chemistry. Understanding of the exciton dynamics in the regime of
strong confinement is important, not only for purely scientific
reasons. The rate of spontaneous emission determines the
statistics of the output of a single photon
source\cite{2000Michler}, the light-intensity of incoherent
sources, consisting of nanocrystal assemblies, such as
LEDs\cite{2002Coe} and the output of coherent sources such as
lasers\cite{2004Wanga}. In addition, the high luminescence
efficiency combined with a narrow homogeneous linewidth make
nanocrystals ideal probes in photonic
studies\cite{2000Michler,2004Lodahl}.

In this Letter, we report on the frequency-dependent decay rate of
excitons in colloidal CdSe and CdTe nanocrystals at room
temperature. We show that the luminescence decay curves, at a
given frequency, are very close to single-exponential. For both
CdSe and CdTe nanocrystals, the decay rate increases with the
emission frequency in a supra-linear way, in contrast to a linear
relation for an ideal two-level exciton system or a cubic relation
for an ideal two-level atom. From complementary calculations based
on tight-binding theory, we conclude that the supra-linear
increase is caused by thermal population of various hole states
with low transition probability (i.e. dark states) that lie close
to the ground state. The frequency dependent rate can thus be
understood on basis of elementary quantum mechanics. The excellent
agreement with the theoretical rates shows that measured rates are
completely determined by radiative decay and highlights the
importance of exciton storage in dark states.

The rate of a spontaneous transition from an excited electron-hole
state $|j\rangle$ to the ground state $|0\rangle$ can be derived
from Fermi's 'golden rule'
\begin{equation}\label{fermi} \Gamma_{j}=\frac{e^{2}}{3
\pi \epsilon_{0} m^{2} \hbar c^{3}} \omega_{j}|\langle
0|\textbf{p}|j\rangle|^{2}
\end{equation}
where $e$ is the electron charge, $\epsilon_{0}$ the permittivity
of free space, $m$ the electron rest mass, $\hbar$ Planck's
constant, $c$ the speed of light, $\omega_{j}$ is the frequency of
the emitted light, and $\langle 0|\textbf{p}|j\rangle$ is the
matrix element of the momentum that is related to the dipolar
matrix element by $\langle
0|\textbf{p}|j\rangle=im\omega_{j}\langle 0|\textbf{r}|j\rangle$
\cite{1996Efros,Delerue}. The size-dependence of $\Gamma_{j}$ will
be determined by the size-dependence of the matrix element
$\langle 0|\textbf{p}|j\rangle$ and emission frequency
$\omega_{j}$. Assuming that the electron-hole states near the
band-extrema can be written as a product of a Bloch function
$\mu(\textbf{r})$ and an envelope function $\phi(\textbf{r})$, it
can be shown that the matrix element of the momentum for an
allowed transition is given by $\langle \mu_{\rm
c}|\textbf{p}|\mu_{\rm v}\rangle$ \cite{1996Efros}, where c and v
denote the conduction and valence bands, respectively. Inter-band
transitions are thus largely determined by the Bloch functions,
which are defined by the crystal lattice only. As a consequence,
$\langle \mu_{\rm c}|\textbf{p}|\mu_{\rm v}\rangle$ does not
depend on the size of the nanocrystals, and the decay rate of an
ideal two-level exciton is expected to be proportional to the
emission frequency; $\Gamma_{j}=constant\cdot\omega_{j}$. This is
in agreement with several theoretical
studies\cite{1984Brus,1987Takagahara,1988Kayanuma}.

Experimental data on the size-dependent strength of the optical
transitions are limited. The extinction coefficient has been
studied for several nanocrystal
suspensions\cite{1993Rajh,1994Vossmeyer,2002Leatherdale,2002Striolo,2003Yu}.
However, the data are not sufficient to show how the squared
dipolar matrix element depends on the size of the nanocrystal.
Direct determination of the radiative lifetime from luminescence
decay curves has proved troublesome due to the fact that the decay
curves are influenced by both non-radiative and radiative
recombination. As a consequence, decay curves are often
multi-exponential\cite{2003Crooker}. This must means that not all
the nanocrystals emit light at the same rate or the decay rate
varies in time \cite{2004Fisher}.
\begin{figure}
\includegraphics[width=0.7\columnwidth]{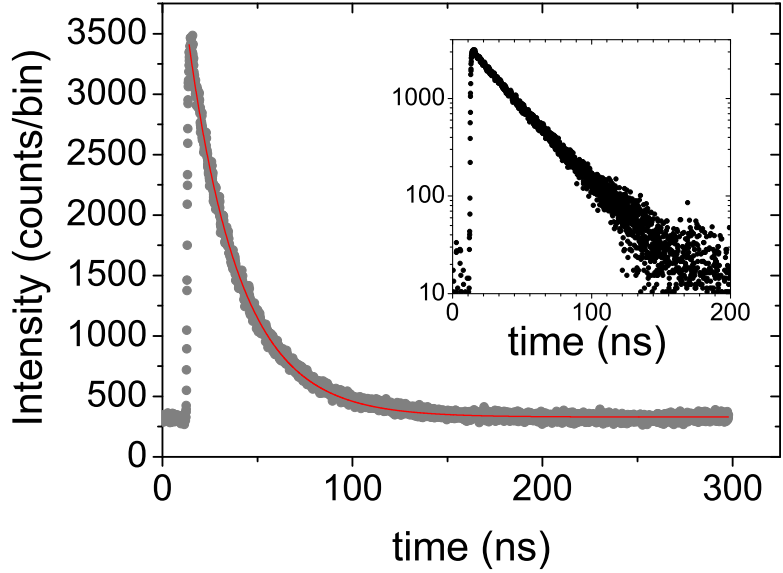}
\caption{\label{Fig1}(Color)Luminescence decay curve of exciton
emission from CdSe nanocrystals at 620 nm $\pm 5$ nm (filled
circles). Single-exponential fit (red curve), with background as
adjustable parameter, yields a rate of 0.037 $\pm$ 0.003 ns$^{-1}$
and a $\chi^{2}_{r}$-value of 1.4. A stretched-exponential model
yields a rate of 0.036 ns$^{-1}$ and a $\beta$-value of 0.99.
Inset: the background corrected decay curve in a semilogarithmic
plot.}
\end{figure}
Encouraged by the availability of suspensions of CdSe and CdTe
nanocrystals\footnote{We have prepared
CdTe\cite{2001Talapin,2003Wuister} and ZnSe[CdSe]\cite{2002Reiss}
nanocrystals with a luminescence quantum yield of 50 $\%$ and
without defect-related emission. Nanocrystal suspensions contain a
collection of differently sized particles, distributed around a
certain average size. The photoluminescence of such a suspension
is inhomogeneously broadened\cite{2003Koberling}. The
size-dependent decay rate can therefore be probed by measurements
at varying frequency in the emission band of the suspension.} with
a high photoluminescence efficiency ($\geq50\%$) we decided to
study the exciton decay rate as a function of the frequency of the
emitted light. The decay curves were obtained by time-correlated
single photon counting. Emission was excited with a Pico Quant
pulsed laser (100 ps, 406 nm) and detection with a monochromator
(0.1 m focal length, 1350 lines/mm grating, blazed at 500 nm)and a
Hamamatsu photomultiplier tube. In Fig.\ref{Fig1} a luminescence
decay curve for a CdSe suspension is shown. The data are
well-described by a single-exponential model, as confirmed by a
goodness of fit $\chi^{2}_{r}$ = 1.4.  A stretched-exponential
model resulted in a stretch parameter $\beta$ = 0.99, very close
to the single-exponential limit of $\beta$ = 1.
\begin{figure}
\includegraphics[width=0.7\columnwidth]{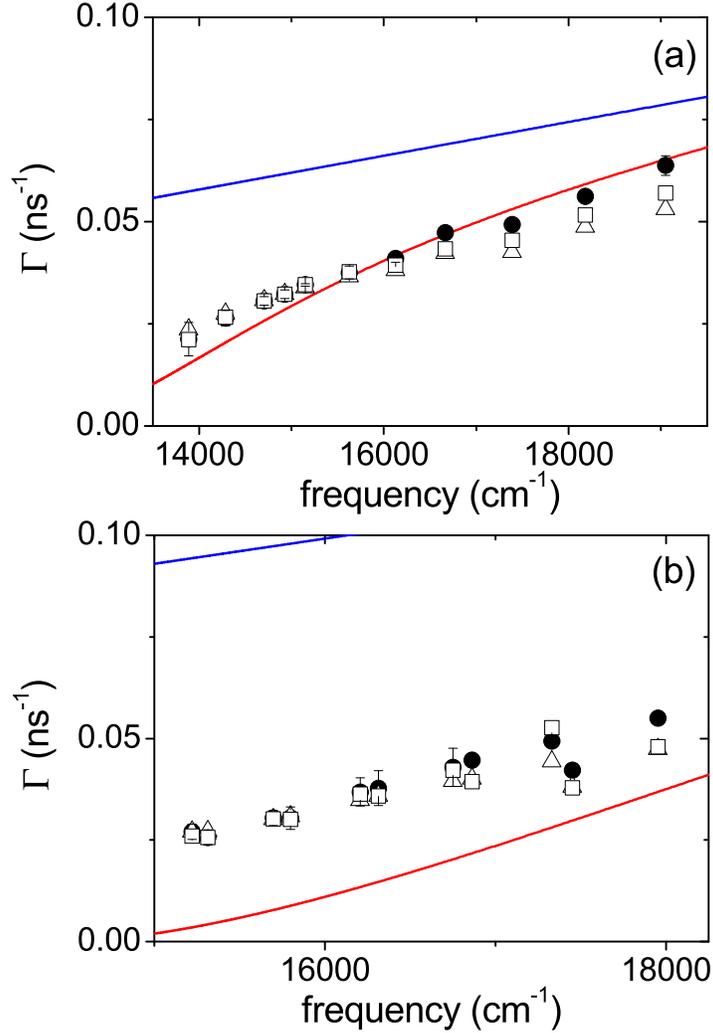}
\caption{\label{Fig2abcd}(Color)Decay rate of emission from CdTe
(a) and CdSe (b) nanocrystals. The decay rate is deduced with
single-exponential fit (filled black dots), average arrival time
(open triangles) and stretched exponential fit (open squares). The
blue lines show the decay rates for a two-level system and pass
through the origin. The red curves show the rate according to Eq.
\ref{gammaradtot} and take dark excitonic states into account.}
\end{figure}

In Fig.\ref{Fig2abcd}, the decay rate is plotted versus the
emission frequency for CdTe and CdSe nanocrystals, respectively.
Data from two CdTe and two CdSe suspensions are plotted in each of
the figures. These data unambiguously show for the first time that
the rate of excitonic decay increases strongly with frequency. The
decay rate was deduced with three different
procedures\footnote{The decay rate was deduced with stretched
exponential\cite{1980Lindsey} $f(t)=\left(\frac{\beta}{t}
\right)(t\Gamma)^\beta\exp(-\Gamma t)^\beta$ where $\beta$ varies
between zero and one. The $\beta$-value illustrates how close the
decay-curve is to single-exponential; $\beta=1$ means that the
decay-curve is single-exponential. The expectation value is
$\langle t\rangle=\frac{1}{\Gamma_{av}}=\int_{0}^{\infty}t
f(t)dt=\frac{1}{(\Gamma\beta)} Gamma[\frac{1}{\beta}]$ where
Gamma[1/$\beta$] is the gamma function. The average time between
the start-pulse and the stop-pulse (i.e. the photon arrival time)
was calculated from the experimental data. For single-exponential
decay this average is equal to the decay time.}; the luminescence
decay curve was fitted with a single-exponential, a stretched
exponential and the average arrival time was calculated. The
$\chi^{2}_{r}$-values and the $\beta$-values were deduced from
single exponential and from stretched exponential fits,
respectively. The $\chi^{2}_{r}$-values are low ($\leq$2) and the
$\beta$-value varies between 0.91 and 1.00. Most of the curves
have a $\beta$-value larger than 0.99. Clearly the decay curves
are nearly single-exponential which means that, at a given
frequency, all the nanocrystals emit light with the same rate.
This forms a first indication that our decay curves are completely
determined by radiative decay\cite{2004Fisher}. We can rule out
the possibility of energy transfer between closely-spaced
nanocrystals\cite{2002Crooker}: first the concentration of
nanocrystals was intentionally kept very low ($<10^{-7}$ mol/l),
second, the results for (sterically stabilized) CdTe nanocrystals
were compared with charge-stabilized nanocrystals, which strongly
repel each other at short distances. Thus, we conclude that the
exciton emission decay rates increase with frequency.

Figure \ref{Fig2abcd} reveals that the experimental decay rates
increase faster with emission frequency than the linear behavior
of a two-level exciton system (blue lines).  Therefore, we must
consider the complex valence band structure of CdTe and CdSe;
several hole-levels are located close to the top of the band. This
leads to other exciton states close to the lowest energy exciton.
Excitons in these states have a much lower probability for
spontaneous decay to the ground state and a high rate of exchange
with the lowest energy exciton. Due to the thermal distribution,
the population of the lowest-energy exciton will be reduced and
thus the rate of radiative recombination will be lowered. If an
infinite lifetime is assumed for the higher-energy excitons, it
can be shown that:
\begin{equation}\label{gammarad}
\Gamma_{\rm rad}=\Gamma_{1}
\left[1+\sum^{N}_{j=2}\exp\left(\frac{-\Delta E_{j}}{kT}   \right)
\right]^{-1}
\end{equation}
where $k$ is Boltzmann's constant, $T$ the temperature,
$\Gamma_{1}$ is the decay rate for the lowest exciton state as
given by Eq.(\ref{fermi}) and $\Delta E_{j}$ is the energy
separation with respect to the ground exciton state. The sum can
be approximated using the density of states of a macroscopic
crystal\cite{Kittel} leading to
\begin{equation}\label{gammaradtot}
\Gamma_{\rm rad}=A \, \omega \left[ 1+ \frac{R^{3}}{6\sqrt\pi}
\left( \frac{2m^{*}kT}{\hbar^{2}} \right)^\frac{3}{2} \right]^{-1}
\end{equation}
where $A$ is a constant, $m^{*}$ is the hole effective mass and
$R$ is the crystal radius, which determines the emission
frequency. Verifying eq. \ref{gammaradtot} by temperature
dependent experiments has proved troublesome due the limited
temperature window in which the quantum efficiency is
constant\cite{2004Wuister}. The effect of thermal occupation of
optically non-active states as a function of crystal size is
caught by the separation between the different exciton levels
($\Delta E_{j}$). For nanocrystals with a larger radius the total
number of states is larger and therefore the separation between
the states is smaller. As a consequence, thermal population of
higher exciton levels is more important in larger nanocrystals.
This means that the radiative decay rate of the large crystals at
low frequency is more reduced than the rate of small crystals at
high frequency. Consequently, the decay rate increases
supra-linearly with frequency, instead of linearly for an ideal
two-level exciton system. The exciton decay rate calculated with
Eq.\ref{gammaradtot}, with $A$ as the only adjustable parameter,
is presented in Fig.\ref{Fig2abcd} (red curves) as a function of
the frequency of the emitted light. The relation between $R$ and
$\omega$ was obtained from tight-binding calculations. Clearly,
the red curves show agreement with our measurements in the case of
CdSe and excellent agreement in the case of CdTe. The supra-linear
dependency of the decay rate can thus be understood on basis of
Fermi's golden rule, if thermally activated dark excitonic states
are accounted for.

\begin{figure}
\includegraphics[width=0.7\columnwidth]{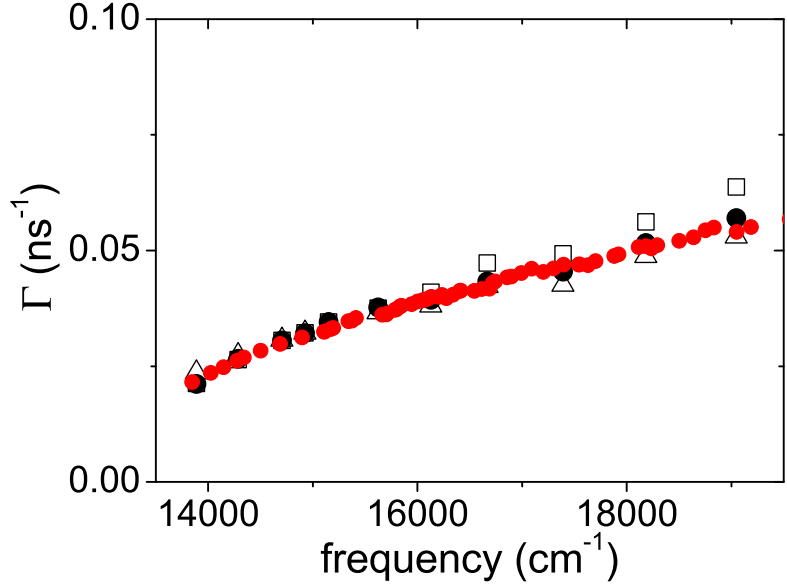}
\caption{\label{Fig3}(Color)Radiative decay rate from
tight-binding calculations (red dots) for CdTe (same experimental
data as in Fig.\ref{Fig2abcd}(a) are plotted.).}
\end{figure}
At a given temperature, the rate of radiative exciton decay
depends on the details of the band structure, determined by the
crystal lattice. In the above considerations several
approximations are used to obtain Eq.(\ref{gammaradtot}) i.e.,
infinite lifetimes for higher exciton states and density of states
of a macroscopic crystal. While this is helpful to get physical
insight, a more detailed calculation, based on tight-binding or
pseudo-potential models, is required to obtain quantitative decay
rates. We have performed tight-binding calculations of the
radiative decay rate for spherical zinc blende CdTe and CdSe
nanocrystals with various sizes\footnote{The tight binding
Hamiltonian matrix is written in a $sp^{3}d^{5}s^{*}$ basis
including spin-orbit coupling and with interactions restricted to
first nearest neighbors. The parameters are obtained by fitting
the experimental effective masses and a reference bulk band
structure calculated using the \emph{ab initio} pseudopotential
code ABINIT in the local density approximation. These parameters
are transferred without change from the bulk situation to the
nanocrystals passivating the surfaces with pseudo-hydrogen atoms.
The matrix elements of ${\bf p}$ are calculated following
Ref.\cite{2004Allan}}. The calculated decay rates for CdTe are in
quantitative agreement with the experimental results over the
accessible frequency range (see red dots in Fig.\ref{Fig3}). This
shows that non-radiative decay has a negligible influence on our
decay curves. In the case of CdSe the tight-binding results show a
supra-linear relation between the radiative decay rate and the
emission frequency (in agreement with the experimental results),
but the absolute value is 75\% too low. To understand the origin
of this discrepancy, we have performed calculations using several
sets of tight-binding parameters. We find that the absolute values
of the emission rates are sensitive to these parameters due to a
subtle coupling between close-spaced hole states. A similar
discussion on the splitting and ordering of levels in the
effective mass approximation can be found in the
literature\cite{1996Richard,1998Efros,1998Fu}. Importantly, we
find that the supra-linear evolution with frequency remains
unaffected for all parameter sets.

The agreement between the supra-linear trend in experiment and in
theory must mean that the experimental decay curves are almost
completely determined by radiative decay. Based on this we argue
that the non-radiative decay rate can be neglected. The
correlation between single-exponential decay curves and a
negligible non-radiative rate is confirmed by results on single
nanocrystals\cite{2002Ebenstein,2004Fisher}. Ref.\cite{2004Fisher}
reports that when a single nanocrystal reveals single-exponential
decay, its non-radiative decay rate is negligible.  For a large
number of nanocrystals with the same diameter of 6.5 nm and
emission frequency $\sim$1.7$\times$10$^4$cm$^{-1}$, a radiative
single-exponential lifetime of 25 ns was observed, in excellent
agreement with our results. In Ref.\cite{2002Ebenstein} it is
shown that a significant fraction of the nanocrystals is
completely dark. This necessarily means that the efficiency of the
emitting nanocrystals is much higher than the efficiency of the
suspension. Therefore, our measured 50\% quantum efficiency can be
explained in simple terms as follows: half of the nanocrystals
have a low emission efficiency, while the other half have
near-unity emission efficiency.  It is the photons from these
latter nanocrystals that mostly contribute to our measured
signals.

We have shown for the first time how the spontaneous emission rate
of semiconductor nanocrystals depends on the frequency of the
emitted light. Comparison with theory shows that spontaneous
emission is considerably attenuated due to occupation of dark
excitonic states. These results may lead to a better understanding
of a number of dynamic effects that are currently studied, such as
exciton dephasing, radiative recombination of nanocrystals in
photonic crystals\cite{2004Lodahl} and F\"orster energy
transfer\cite{1996Kagan,2002Crooker}.

We thank P. Vergeer and J. J. Kelly for fruitful discussions. This
work was supported by the Stichting Fundamenteel Onderzoek der
Materie (FOM) and Chemische Wetenschappen (CW) with financial aid
from the Nederlandse Organisatie voor Wetenschappelijk Onderzoek
(NWO).

\end{document}